\documentclass[usenatbib]{mn2e}
\usepackage{graphicx}

\def\apj{\rm ApJ}
\def\apjl{\rm ApJL}
\def\apjs{\rm ApJS}
\def\aj{\rm AJ}
\def\mnras{\rm MNRAS}
\def\nat{\rm Nature}
\def\pasj{\rm PASJ}
\def\pasp{\rm PASP}

\def\aap{\rm AAP}
\def\araa{\rm ARA\&A}

\def\gax{\mathrel{\raise.3ex\hbox{$>$}\mkern-14mu\lower0.6ex\hbox{$\sim$}}}
\def\lax{\mathrel{\raise.3ex\hbox{$<$}\mkern-14mu\lower0.6ex\hbox{$\sim$}}}
\def\gtorder{\mathrel{\raise.3ex\hbox{$>$}\mkern-14mu
             \lower0.6ex\hbox{$\sim$}}}
\def\ltorder{\mathrel{\raise.3ex\hbox{$<$}\mkern-14mu
             \lower0.6ex\hbox{$\sim$}}}

\begin{document}

\title [Cas~A and the Crab Were Not Binaries]
   {Cas~A and the Crab Were Not Stellar Binaries At Death}

\author[C.~S. Kochanek et al.]{ 
    C.~S. Kochanek$^{1,2}$, 
    \\
  $^{1}$ Department of Astronomy, The Ohio State University, 140 West 18th Avenue, Columbus OH 43210 \\
  $^{2}$ Center for Cosmology and AstroParticle Physics, The Ohio State University,
    191 W. Woodruff Avenue, Columbus OH 43210 \\
   }

\maketitle

\begin{abstract}
The majority of massive stars are in binaries, which implies that many
core collapse supernovae (ccSNe) should be binaries at the time of the explosion.  Here we
show that the three most recent, local (visual) SNe (the Crab, Cas~A and SN~1987A) 
were not binaries, with limits on the initial mass ratios of $q=M_2/M_1 \ltorder 0.1$.   
No quantitative limits have previously been set for Cas~A and the Crab, while for SN~1987A
we merely updated existing limits in view of new estimates of the dust content.
The lack of stellar companions to these three ccSNe implies a 90\% confidence upper limit on the 
$q\gtorder 0.1$ binary fraction at death of $f_b < 44\%$.  In a passively evolving binary model 
(meaning no binary interactions), with a flat mass ratio distribution and a 
Salpeter IMF, the resulting 90\% confidence upper limit on the initial binary fraction of 
$F < 63\%$ is in considerable tension with observed massive binary statistics.
Allowing a significant fraction $f_M \simeq 25\%$ of stellar binaries to merge
reduces the tension, with $F < 63(1-f_M)^{-1}\% \simeq 81\%$, but allowing for 
the significant fraction in higher order systems (triples, etc.) reintroduces 
the tension.  That Cas~A was 
not a stellar binary at death also shows that a massive binary companion is not 
necessary for producing a Type~IIb SNe.  
Much larger surveys for binary companions to Galactic SNe will become feasible with
the release of the full Gaia proper motion and parallax catalogs, providing a powerful
probe of the statistics of such binaries and their role in massive star evolution,
neutron star velocity distributions and runaway stars.
\end{abstract}


\begin{figure}
\centering
\includegraphics[width=0.45\textwidth]{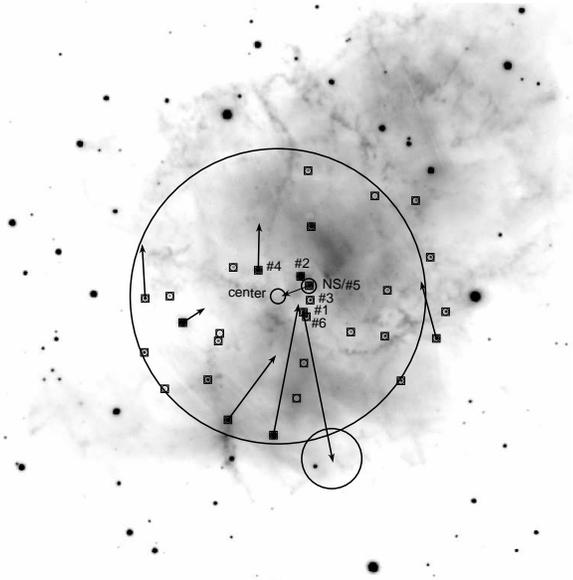}
\caption{ Coadded grizy PS1 image of the Crab.  The position of the geometric center of the remnant
  (``center'') and the neutron star are indicated by 3\farcs0 radius green circles.  The larger green
  circle shows the region within 60\farcs0 of the center.  The 30 stars within 60\farcs0 of either
  the center or the NS are marked, and the six closest to the center are numbered in order of
  their distance from the center.  The NS is star \#5.  The arrows show the predicted positions
  of sources with proper motions at the time of the SN.  As expected, the predicted position of 
  the pulsar is close to the center of the SNR.  Seven stars have proper motions in NOMAD with
  uncertainties in their back-projected positions of approximately 12\farcs0 as shown by the  
  circle at the head of one of the proper motion vectors.  At a distance of $2.0$~kpc, a star
  will have moved $10\farcs1(v/100~\hbox{km/s})$ since the SN, so the 60\farcs0 search radius 
  corresponds to a velocity of roughly 600~km/s.  
  }
\label{fig:image1}
\end{figure}

\begin{figure}
\centering
\includegraphics[width=0.45\textwidth]{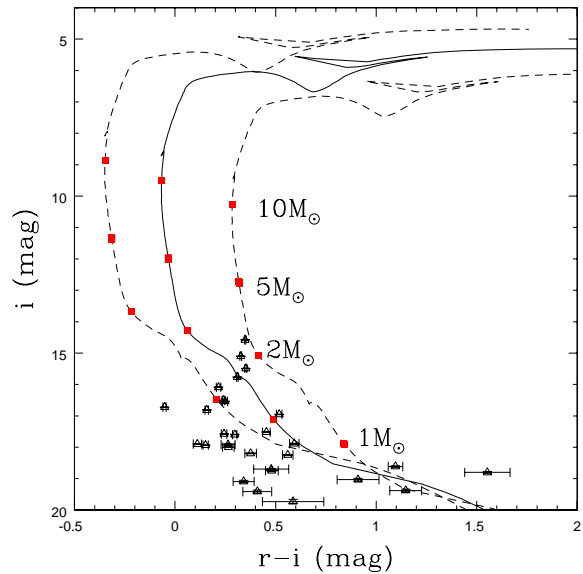}
\caption{ The r/r$-$i CMD of the stars within 60\farcs0 of the center of the Crab SNR or NS. 
  The solid curve shows the PARSEC (\protect\citealt{Bressan2012})
  isochrone for Solar metallicity stars with an age of $10^{7.3}$~years at a distance of $2$~kpc and with
  an extinction of $E(B-V)=0.4$~mag.  The dashed curves show
  the effect of reducing (raising) the extinction to $E(B-V)=0$~mag ($0.9$~mag).  Uncertainties in the
  distance modulus are much less important, corresponding to vertical shifts of $\pm 0.5$~mag.
  Red filled squares on the isochrones mark stars with masses of $1$, $2$, $5$ and $10 M_\odot$.
  }
\label{fig:cmd1}
\end{figure}

\begin{figure}
\centering
\includegraphics[width=0.45\textwidth]{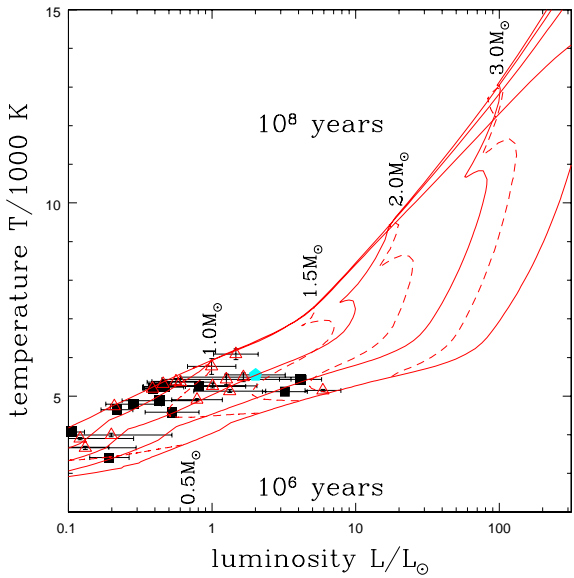}
\caption{ The luminosities and temperatures of the stars if at the distance of the Crab and
  constrained by the extinction prior.  Filled black squares mark the stars that could be
  at the distance of the Crab ($\chi_2^2 < \chi^2_0 + 4$) and an association is not ruled 
  out by the available 
  proper motions.  Open red triangles are for stars that either cannot lie at the
  distance of the Crab ($\chi_2^2 > \chi^2_0 + 4$) or have a proper motion inconsistent
  with an association.  The pulsar, fit as a star, is indicated by the filled, cyan
  pentagon.  The solid lines show 
  isochrones with ages of $10^6$, $10^{6.5}$, $10^7$, $10^{7.5}$ and $10^8$~years
  while the dashed lines show the tracks for $0.5$, $1.0$, $1.5$, $2.0$ and $3.0 M_\odot$
  stars over this range of times.
  }
\label{fig:lum1}
\end{figure}

\begin{figure}
\centering
\includegraphics[width=0.45\textwidth]{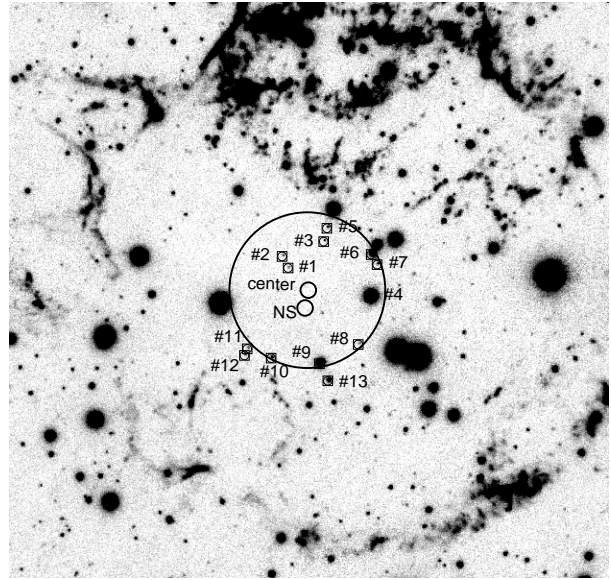}
\caption{ Coadded grizy PS1 image of Cas~A.  The position of the geometric center of the remnant
  (``center'') and the neutron star are indicated by 3\farcs0 radius circles.  The larger 
  circle shows the region within 30\farcs0 of the center.  The 13 PS1 stars 
  lying within 30\farcs0 of either the geometric center or the neutron star are
  marked and labeled in order of their distance from the center.  None of the stars have proper
  motion measurements in NOMAD.  Stars \#4, \#9 and \#13 have proper motions in HSOY
  but the shift in position to the time of the SN is too small to display. 
  At a distance of $3.4$~kpc, a star will have moved 
  $2\farcs1(v/100~\hbox{km/s}) $ since the SN, so the 30\farcs0 search radius 
  corresponds to a velocity of roughly 1500~km/s
  }
\label{fig:image2}
\end{figure}

\begin{figure}
\centering
\includegraphics[width=0.45\textwidth]{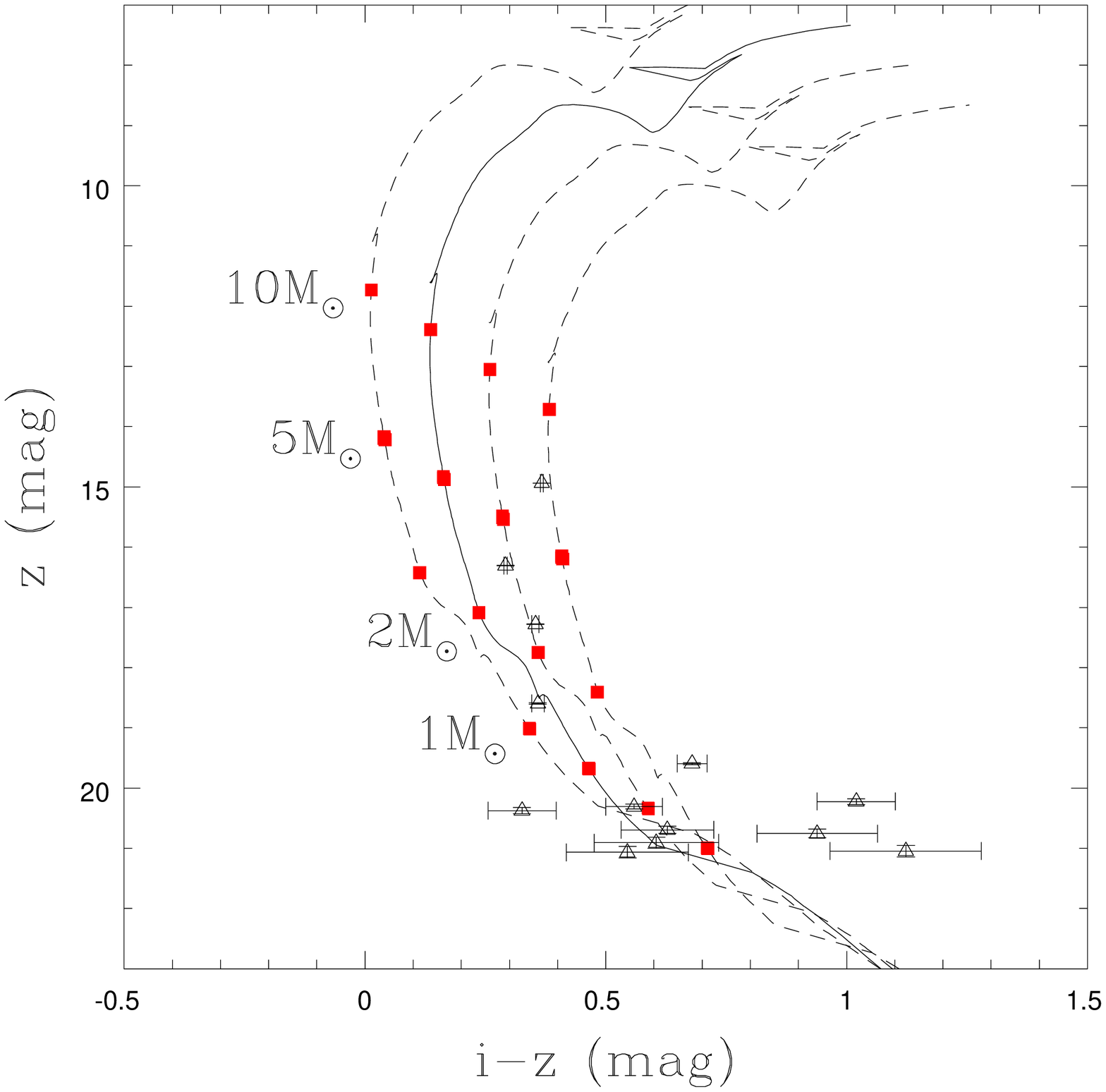}
\caption{ The z/i$-$z CMD of the stars within 30\farcs0 of the center of the Cas~A SNR or the NS.
  The solid curve shows the PARSEC (\protect\citealt{Bressan2012}) isochrone for Solar metallicity 
  stars with an age of $10^{7.3}$~years at a distance of $3.4$~kpc and with an extinction of 
  $E(B-V)=1.6$~mag ($A_V=5.0$).  The dashed curves show the effect of reducing the extinction
  to $A_V=3.4$ or raising it to $6.5$ or $8.0$~mag. Uncertainties in the
  distance modulus are much less important, corresponding to vertical shifts of $\pm 0.2$~mag.
  Red filled squares on the isochrones mark stars with masses of $1$, $2$, $5$ and $10 M_\odot$.
  }
\label{fig:cmd2}
\end{figure}

\begin{figure}
\centering
\includegraphics[width=0.45\textwidth]{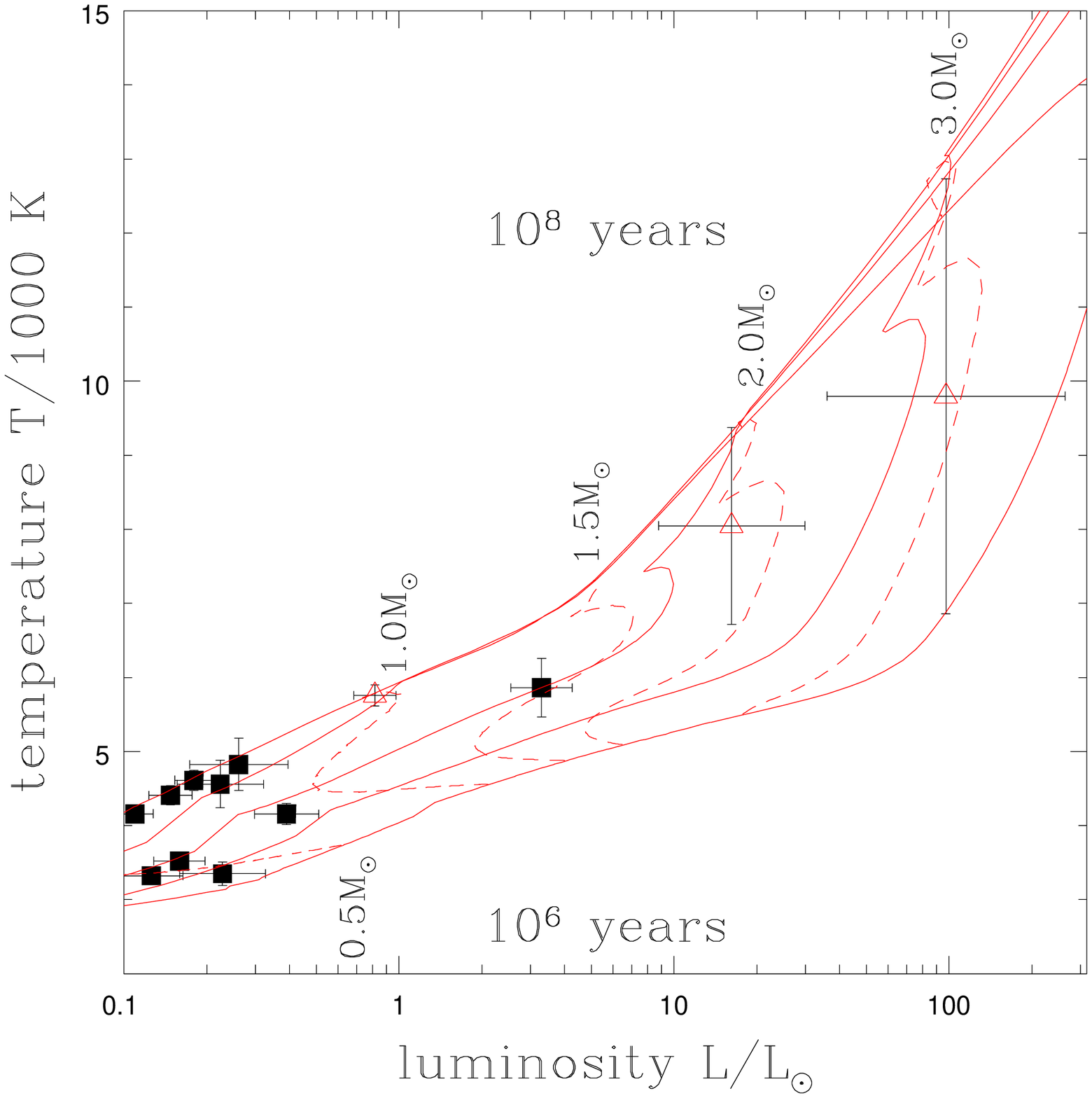}
\caption{ The luminosities and temperatures of the stars if at the distance of the Cas A and
  constrained by the extinction prior.  Filled black squares mark the stars that could be
  at the distance of Cas A ($\chi_2^2 < \chi^2_0 + 4$) and not ruled out by proper motions.
  Open red triangles mark stars that cannot be at the distance of the Crab 
  ($\chi_2^2 > \chi^2_0 + 4$) or an association is ruled out by the proper motions.
  In practice, the only star which is inconsistent with the distance (\#13) is also
  ruled out by its proper motion.  The solid lines show
  isochrones with ages of $10^6$, $10^{6.5}$, $10^7$, $10^{7.5}$ and $10^8$~years
  while the dashed lines show the tracks for $0.5$, $1.0$, $1.5$, $2.0$ and $3.0 M_\odot$
  stars over this range of times.
  }
\label{fig:lum2}
\end{figure}

\section{Introduction}

A large fraction of massive stars appear to be in binaries (see the reviews
by \citealt{Duchene2013} and \citealt{Moe2016}). \cite{Kobulnicky2014}
estimate that 55\% are in binaries with $P<5000$~days and mass ratios of $0.2 < q < 1$,
while \cite{Sana2012} estimate that 69\% are in binaries and that two-thirds of these
will undergo some form of interaction.  \cite{Moe2016} find that
only $16\pm8$\% of the $9$-$16M_\odot$ stars that will dominate the SN 
rate are single, and that they
have an average multiplicity (companions per primary) of $1.6\pm0.2$.

Mass transfer, mass loss and mergers then
significantly modify the subsequent evolution of the system
(e.g., \citealt{Eldridge2008}, \citealt{Sana2012}). 
This will, in turn, modify the properties of any
resulting supernovae (SNe) over the expectations for isolated stars.  For 
example, the numbers of stripped Type~Ibc SNe and the limits on
their progenitor stars both suggest that many are stripped through binary mass
transfer rather than simply wind (or other) mass loss (e.g., \citealt{DeDonder1998},
\citealt{Eldridge2008},
\citealt{Smith2011}, \citealt{Eldridge2013}).  There are many theoretical
studies exploring the stripped Type~IIb, Ib and Ic SNe in the context
of binary evolution models (e.g., \citealt{Yoon2010}, \citealt{Yoon2012}, 
\citealt{Claeys2011}, \citealt{Dessart2012}, \citealt{Benvenuto2013}, \citealt{Kim2015}, \citealt{Yoon2017}),
as well as models for the effects of binary evolution on electron capture
SNe (e.g., \citealt{Moriya2016}).  

Discussions of the binary companions to local core collapse SNe have largely 
focused on understanding runaway B stars (e.g., \citealt{Blaauw1961}, \citealt{Gies1986}, 
\citealt{Hoogerwerf2001}, \citealt{Tetzlaff2011}) and the 
contribution of binary disruption to the velocities of neutron 
stars (NS) (e.g., \citealt{Gunn1970}, 
\citealt{Iben1996}, \citealt{Cordes1998}, \citealt{Faucher2006}). \cite{vandenBergh1980} seems
to have been the first to search supernova remnants (SNRs) for runaway
stars by looking for a statistical excess of O stars close to the 
centers of 17 SNRs and finding none.  \cite{Guseinov2005} examined 48
SNRs for O or B stars using simple color, magnitude and proper motion
selection cuts to produce a list of candidates based on the 
USNO A2 catalog (\citealt{Monet1998}). None of these systems have
been investigated in any quantitative detail.
\cite{Dincel2015} identify a good candidate in the 
$\sim 3 \times 10^4$~year old SNR~S147 containing PSR~J0538+2817 
and argue that it was also likely to have been an interacting binary.  
Considerably more effort has been devoted to searching for single
degenerate companions to Type~Ia SN (e.g., \citealt{Schweizer1980}  
\citealt{Ruiz2004}, \citealt{Ihara2007}, \citealt{Gonzalez2012}, 
\citealt{Schaefer2012}). 
 
Searches for binary companions to core-collapse SNe in external galaxies
are more challenging because the companion is generally significantly 
fainter than the progenitor (see \citealt{Kochanek2009}).  
The Type~IIb SN~1993J is probably the best case (\citealt{Maund2004}, \citealt{Fox2014}),
while the existence of a companion to the Type~IIb SN~2011dh is debated, 
with \cite{Folatelli2014} arguing for a detection and \cite{Maund2015} arguing 
that the flux may be dominated by late time emission from the SN.  There is some
evidence of a blue companion for the Type~IIb SNe~2001ig (\citealt{Ryder2006})
and SN~2008ax (\cite{Crockett2008}.  There are limits on the
existence companions to the Type Ic SNe~1994I (\citealt{VanDyk2016})
and SN~2002ap  (\citealt{Crockett2007}) and the Type~IIP SNe~1987A
(\citealt{Graves2005}), SN~2005cs (\citealt{Maund2005}, \citealt{Li2006}),
and SN~2008bk (\citealt{Mattila2008}).  All of these limits assume that
the SNe made little dust, an issue we discuss further in \cite{Kochanek2017}
and consider in more detail for SN~1987A below.

\cite{Kochanek2009} examined the statistical properties expected for
surviving binary companions to SNe assuming passively evolving systems
(i.e. no binary interactions).  As already noted, the companions are
generally significantly fainter than the exploding star, although this
is frequently not the case for stripped SN progenitors -- for Type~Ibc
SNe, it should not be surprising to find that the binary companion 
is more visually luminous than the SN progenitor. This point is of
considerable importance for the one candidate Type~Ib progenitor
iPTF13bvn (\citealt{Cao2013},
\citealt{Groh2013}, \citealt{Bersten2014}, \citealt{Fremling2014},
\citealt{Eldridge2015}, \citealt{Eldridge2016},
\citealt{Folatelli2016}).  If the initial
binary fraction is $F$, then the fraction of passively evolving binaries
that are in stellar binaries at death is
\begin{equation}
    f_b = { F \over 1 + F f_q }
   \quad\hbox{where}\quad f_q = \int_{q_{min}}^{q_{max}} q^{x-1} P(q) dq,
   \label{eqn:bfrac}
\end{equation} 
$x \simeq 2.35$ is the slope of the initial mass function (IMF),
$q_{min} \leq q = M_2/M_1 \leq q_{max} \leq 1$ is the mass ratio and
$P(q)$ with $\int dq P(q) \equiv 1$ is the distribution of mass 
ratios.  For a Salpeter IMF and a flat $P(q)$ distribution extending
over $0 \leq q \leq 1$, $f_q=0.426$ and the fraction of SNe in 
stellar binaries at death is 23\%, 41\%, 57\% and 70\% for initial
binary fractions of $F=25\%$, 50\%, 75\% and 100\%, respectively.
Essentially, only the explosions of primaries occur in stellar
binaries, so the fraction of SNe in stellar binaries is less than
the initial fraction of binaries because some of the SNe are the
explosions of secondaries.  Binary evolution, particularly stellar
mergers, then adds further complications,
as does the prevalence of triples and other higher order systems.

For a supernova of a given age, we need an estimate of the radius
inside the SNR that needs to be searched.  \cite{Guseinov2005}
simply used a fixed $1/6$ of the diameter of the remnant as
cataloged by \cite{Green2014} (for the most recent version).
Observed runaway stars have typical velocities of $50$~km/s
or less (e.g., \citealt{Tetzlaff2011}) and theoretical studies
find that it is difficult for binaries to produce velocities
of more than a few 100~km/s (e.g., \citealt{Cordes1998}, \citealt{Eldridge2011}).
This has a simple explanation in terms of stellar structure because
the maximum (circular) orbital velocity of a secondary star is limited by
\begin{equation}
    v_2^2 < { G M_1 \over 1 + q } \left[ { 4 \pi \sigma T_1^4 \over L_1 }\right]^{1/2}
          \left[ 1 + \left( { L_2 \over L_1} \right)^{1/2} { T_1^2 \over T_2^2 } \right]^{-1}
\end{equation}
where the mass ratio $q=M_2/M_1$ could be $>1$ here, 
the semi-major axis is set to the sum of the two stellar
radii, and we have expressed the radii in terms of the stellar luminosities and
effective temperatures.  The highest possible 
companion velocity is achieved for a low mass  ($q \rightarrow 0$) and 
low luminosity $L_2/L_1 \rightarrow 0$ companion.  This
allows the simple upper limit on the companion's velocity of
\begin{equation}
    v < 50 M_{10}^{1/2} T_{3.5} L_{4.7}^{-1/4}~\hbox{km/s}
    \label{eqn:vlimit}
\end{equation}
where the scalings of $M_1 = 10 M_{10} M_\odot$, $T_1 = 10^{3.5} T_{3.5}$~K and 
$L_1 = 10^{4.7} L_{4.7} L_\odot$ are chosen to match the end point of a
$10^{7.4}$~year PARSEC (\citealt{Bressan2012}) stellar isochrone.  
The scaling with mass and luminosity is very weak\footnote{
For $L \propto M^x$, the velocity limit scales with primary mass as 
$v_1 \propto M^{1/2-x/4}$. For $x=3$ ($x=2$), this becomes 
$v_1 \propto M_1^{-1/4}$ ($v_1 \propto M_1^0$).}, so the only important
variable is the temperature of the primary.  For the typical red 
supergiant progenitors of Type~II SNe (see \citealt{Smartt2009}), we
expect very low velocities, $v < 50$~km/s.  In the rare cases like SN~1987A 
where the primary is a blue supergiant at the time of explosion, the maximum 
companion velocity is still $v \ltorder 300$~km/s.    

Companions to stripped, Type~Ibc SNe can have higher velocities 
because of the very high progenitor temperatures.  For these systems,
the finite size of the secondary is important because the radius
of the primary is $ \sim R_\odot$, and the companion velocities
can in theory reach $\sim 10^3$~km/s.  This is only 
true if the system was an interacting binary because the orbit of
the secondary must also shrink to be far smaller than even the
initial size the primary.  Such tightly bound binaries are less likely to
be disrupted because the primary mass has to have been greatly
reduced by mass loss and the orbital binding energy is larger
than typical NS kick velocities (e.g., \citealt{Cordes1998}). 
Theoretically, \cite{Eldridge2011}, using binary population synthesis
models that included such evolutionary paths, found that velocities 
above $300$~km/s were very rare.

Here we consider the three most recent visually observed, Local Group,
core collapse SNe: the Crab, Cas~A and SN~1987A.  For Galactic SN,
the Crab and Cas~A have several advantages.  Their youth means that
the search areas are small, and the fact that they were visible by
eye means that they have modest extinctions and, by extension,
lie in regions with relatively low stellar densities for the Galaxy.  We were
unable to find any quantitative discussions of searches for binary
companions to these two systems, but \cite{Guseinov2005} report
no candidates in their qualitative survey of 48 SNRs.  The Crab pulsar is 
observed in the optical/near-IR, but it is emission due to the pulsar 
and not from a surviving binary (e.g., \citealt{Sandberg2009}, \citealt{Scott2003}).  
There have been a series of unsuccessful searches for an optical/near-IR counterpart 
to the NS in Cas~A which rule out any bound system even at the level
of a $M_2 \simeq 0.1 M_\odot$ dwarf companion (e.g., \citealt{vandenBergh1986}, 
\citealt{Kaplan2001}, \citealt{Ryan2001}, \citealt{Fesen2006}). These
studies also implicitly set strong limits on any unbound system, but
the topic is never discussed in these papers.  \cite{Graves2005} set very
strong limits on the existence of a binary companion to SN~1987A but
assumed there was very little dust obscuration created by the SN. 
More recent studies have shown that SN~1987A formed far more dust
than assumed by \cite{Graves2005} and that it is
concentrated towards the center of the remnant
(\citealt{Matsuura2011}, \citealt{Indebetouw2014}, \citealt{Matsuura2015}),
making it necessary to revisit these limits.

The Crab was almost certainly a Type~II
SNe due to the presence of significant amounts of hydrogen. 
However, the SNR appears to contain too little mass or energy
for it to have been a normal Type~II SN, suggesting it may have
been an electron capture SN (see the review by \cite{Hester2008} 
or the recent discussion by \cite{Smith2013}). The binary models
of \cite{Moriya2016} would be one way of having such a low ejecta mass.  
Cas~A is known to be a Type~IIb thanks to spectra of light echoes from the
SN (\citealt{Krause2008}, \citealt{Rest2008}, \citealt{Rest2011}, 
\citealt{Finn2016}).  Single star evolution models generally 
have difficulty producing Type~IIb SNe (e.g., \citealt{Young2006}
for Cas~A in particular, \citealt{Podsiadlowski1993}, \citealt{Woosley1994},
\citealt{Claeys2011}, \citealt{Dessart2012}, 
\citealt{Benvenuto2013}, more generally).  
While SN~1987A was a Type~II SN, the
progenitor was also a blue rather than a red supergiant (see
the review by \citealt{Arnett1989}).  Several models have invoked
binary interactions, possibly with a final merger, to explain
either the structure of the star or the surrounding winds
(e.g., \citealt{Podsiadlowski1989}, \citealt{Podsiadlowski1992}, 
\citealt{Blondin1993}, \citealt{Morris2009}). 

Here we take advantage of the recently released PS1 survey data
(PS1, hereafter, \citealt{Chambers2016}) along with their associated three-dimensional 
maps of dust in the Galaxy (\citealt{Green2015}) to examine stars
near the centers of the Crab and Cas~A quantitatively.  For consistency
with the dust maps, we use the extinction coefficients $A_\lambda$
from \cite{Schlafly2011}.
We also use, where available, 
the NOMAD (\citealt{Zacharias2005}) or HSOY (\citealt{Altmann2017}) proper motions.  We fit the 
photometry of stars near the center of the SNR using Solar metallicity 
stars drawn from the PARSEC isochrones (\citealt{Bressan2012}). For
the coarse luminosity estimates we require, the effects of metallicity
on stellar colors should not be very important.   For SN~1987A we
simply examine the consequences of the new dust observations.
In sections \S2-4 we discuss the Crab, Cas~A and SN~1987A in turn,
and we discuss the implications of the results in \S5.

\section{The Crab (SN~1054) }

Figure~\ref{fig:image1} shows the co-added grizy PS1 image of a roughly 
2~arcmin region around the center of the Crab SNR.  
We adopt an age of 960 years and, following \cite{Kaplan2008}, a distance of 
$2.0 \pm 0.5$~kpc or $\mu = 11.51\pm 0.54$ as a distance 
modulus.  Using the \cite{Green2015} dust distribution
for the line of sight towards the center of the SNR, the extinction
is roughly $E(B-V) \simeq 0.4$~mag.  \cite{Green2015} estimate
that the dust distribution is well-defined out to a distance modulus
of $14.2$ which is well beyond the distance to the Crab. This 
extinction estimate agrees well with other
determinations (e.g., \citealt{Wu1981}, \citealt{Blair1992}).
We define the (J2000) center of the SNR as (05:34:32.84, 22:00:48.0) from 
\cite{Nugent1998} and the position of the pulsar as 
(05:34:31.9, 22:00:52.1).  
\cite{Kaplan2008} measured the proper motion of the pulsar, and its 
estimated position at the time of the SN agrees very closely with 
the estimated center of the SNR.  The center of the SNR and the
position of the NS at present and in 1054 are both marked in 
Figure~\ref{fig:image1}. 

For a distance of $2.0$~kpc and an age of 960~years, a star with
velocity $v = 100 v_2$~km/s has moved $10\farcs1 v_2$.  We 
selected stars within 1~arcmin of either the center of the SNR
or the NS, corresponding to a velocity limit of approximately
$600$~km/s.  Since the proper motion of the NS only corresponds to
$v \simeq 100$~km/s and the Crab is believed to have been a 
Type~II SN (even if peculiar) for which Equation~\ref{eqn:vlimit}
applies, only sources within roughly 10\farcs0 of the center
can plausibly be surviving secondaries.  

The PS1 catalog for these 1~arcmin regions contains 171 sources, most of which 
are spurious detections created by the bright nebular emission or faint 
sources in the wings of the brighter point sources.  When we went to
fit models to the spectral energy distributions (SEDs) of the actual stars, 
we frequently found no good fits ($\chi^2 > 100$) for any stellar model even though
we were using the PS1 PSF magnitudes.  This is very different from the
case of Cas~A (next section), where we almost always found very good model fits.
Presumably this is because the PS1 photometry pipeline was never
intended for photometry of stars in a nebular emission region like
the center of the Crab nebula.  
To remedy this issue, we used {\tt Sextractor} (\citealt{Bertin1996})
to identify sources on the roughly 2~arcmin square co-added PS1 image 
and then used {\tt IRAF} aperture photometry with a 3 pixel (0\farcs75) aperture radius and
a 6 pixel to 10 pixel radius sky aperture.  This larger $2'$ area includes 
many stars beyond the brightest nebular regions. We then matched the aperture
results to the PS1 photometry and computed the necessary photometric
offsets as the median offset after clipping outliers.  In the end
we had 30 stars to consider, labeled in order of distance from the
center of the SNR, including the NS as star \#5.  
Table~\ref{tab:crab1} provides the positions and
grizy aperture magnitudes of these 30 stars along with their distances
from the center of the SNR and the NS.    

There are, in fact, no sources closer than $12\farcs0$ to the center
of the SNR.  Since there is so little ambiguity in the center given
its close (arcsecond) agreement with the proper motion of the pulsar,
and even a $10\farcs0$ search radius is already generous for the
companion of a Type~II SN, we can immediately rule out the existence
of a binary companion with $M_2 \gtorder M_\odot$ simply from the
structure of the CMD.  This could be strengthened further using
the still deeper HST data.
There are six sources between $12\farcs0$ and $15\farcs0$, where
the pulsar is source \#5, and then the next source is 21\farcs7
from the center.  
We number the sources by
their distance from the center of the SNR and have labeled only
the six closest sources in Figure~\ref{fig:image1}.

Figure~\ref{fig:cmd1} shows the r/r$-$i color magnitude diagram (CMD) of
these sources.  
Superposed are the PARSEC isochrones for Solar metallicity stars with
an age of $10^{7.3}$~years, roughly corresponding to the epoch at which
$12 M_\odot$ stars would die, along with the effects of changing the
extinction estimates to $E(B-V)=0$ or $0.9$~mag.  The distance uncertainties
are much less important, since they only correspond to shifting the
isochrone vertically by $\pm 0.5$~mag.  Like \cite{Guseinov2005}, we 
find no plausible candidates for a former binary companion.    

Seven of the stars (\#1, 4, 13, 20, 21, 22 and 28) have proper motions 
in the NOMAD catalog, and their predicted positions at the time of the 
SN are indicated by the arrows in Figure~\ref{fig:image1}.  The 
uncertainties in the proper motions lead to a position uncertainty of 
approximately 12\farcs0 after 960 years, as indicated by the circle
at the end of one of the proper motion vectors.  Only the distant star \#22
has a proper motion consistent with the position of the SN, but it
would also have to be moving at almost $600$~km/s. The two
closer stars with proper motions, \#1 and \#4, are moving in the
wrong direction to be associated with the SN.

To formalize the qualitative impression, we drew a sample of roughly $10^5$
stars from the PARSEC isochrones, uniformly sampling in age from $10^6$ to
$10^{10}$~years in increments of $0.01$~dex.  We fit the PARSEC estimates
of the absolute PS1 magnitudes to all the candidate sources to
estimate the distance and extinction.  We considered four fits to each
of the candidate sources for each of the $\sim 10^5$ stellar models.  First,
we fit for the extinction $E(B-V)$ and distance modulus $\mu$ with no 
constraints on either variable.  Given magnitudes $m_i$ with uncertainties $\sigma_i$
and a model star with absolute magnitudes $M_i$, the fit statistic is
\begin{equation}
     \chi^2_0 = \sum_i \left( m_i - M_i - \mu - R_i E(B-V) \right)^2 \sigma_i^{-2}.
\end{equation}
Second, we repeated the fits constrained by the estimated
distance modulus $\mu_0 \pm \sigma_\mu$ to the Crab, 
$\chi^2_1 = \chi^2_0 + (\mu - \mu_0)^2/\sigma_\mu^2$.  
For the last fit, we added a prior on the extinction based on the 
\cite{Green2015} estimates of the extinction as a function of distance 
modulus to the fit constrained by the distance modulus.
 We used the variance of their $20$ alternate extinction realizations as
an estimate of the uncertainty in the extinction at any given distance
to give $E(\mu) \pm \sigma_{E(\mu)}$.  
We simply added this prior to the results of the previous fits to
get $\chi_2^2 = \chi^2_1 + (E(B-V)-E(\mu))^2/\sigma_{E(\mu)}^2$.  
We did 
not repeat the fits with the (non-linear) constraint on the extinction 
as a function of distance -- the broad range of the input models and the
simplicity of the conclusions makes this complication unnecessary. We
assumed uncertainties that are the larger of the photometric uncertainties 
and 0.05~mag.  The minimum uncertainty is included to compensate for
modest systematic errors (e.g., extinction law, photometric systems, 
calibrations, metallicity).  

These fits are carried out for each of the $\sim 10^5$ stellar models,
and the best fits are reported in Table~\ref{tab:crab2}. The table contains
the goodness of fit for the best stellar model with no priors ($\chi^2_0$), 
with a prior on the distance ($\chi^2_1$), and with a prior on both the 
distance and the extinction ($\chi^2_2$).  In practice, there are 
reasonable fits for a range of stellar models, which we can characterize
with quasi-Bayesian averages such as     
\begin{equation}
  \langle \log L \rangle = 
    \left[ \sum e^{-\chi'^2/2} \log L \right] 
    \left[ \sum e^{-\chi'^2/2} \right]^{-1} 
\end{equation}
for the luminosity.  In these averages,
$\chi'^2$ means that we have renormalized the $\chi^2$ values
so that the best fit has a $\chi^2$ per degree of freedom of unity
if the raw $\chi^2$ is larger than the number of degrees of freedom.
This has the effect of broadening the uncertainties for sources
which are less well fit.  The same average is carried out for
$\log T$. The allowed spreads of the luminosity and temperatures 
about the averages are estimated by the probability weighted 
dispersion of the solutions about the averages,
\begin{equation}
  \sigma_{\log L}^2  = 
    \left[ \sum e^{-\chi'^2/2} (\log L - \langle \log L \rangle)^2 \right]
    \left[ \sum e^{-\chi'^2/2} \right]^{-1}.
\end{equation}
Table~\ref{tab:crab2} reports these estimates of the luminosities
and temperatures for the models with a prior on the distance but
not on the extinction ($\chi^2_1$) and the model with priors on
both the extinction and the distance ($\chi^2_2$).

As is typical of trying to estimate stellar distances using only
photometry, it is nearly impossible to do so without additional
constraints.  Very few of the stars cannot be placed at the 
distance of the Crab if there is no constraint on the extinction
(5 of 30 have $\chi^2_1 > \chi^2_0 + 4$).  With the addition of
the constraint on the extinction at any given distance included,
many fewer have solutions consistent with the distance 
(15 of 30 have $\chi^2_2 > \chi^2_0 + 4$), but that still 
leaves many that are consistent with both priors.  Curiously,
the non-thermal emission of the NS (\#5) can be well-modeled
by a stellar SED.  
 
Figure~\ref{fig:lum1} shows the luminosities and temperatures the
stars would have at the distance of the Crab and with the extinction
prior ($\chi^2_2$).  The distribution looks similar without the extinction prior,
but the uncertainties, particularly the temperature uncertainties,
become larger (see Table~\ref{tab:crab2}). If any of these stars
are at the distance of the Crab, none of them are either 
luminous or massive.
Moreover, many of the more luminous stars are also the ones
with proper motions, almost all of which are inconsistent with
an association to SN~1054.  More importantly, an actual 
companion to the Crab SN would have to be closer to the 
center of the SNR where no stars of similar magnitudes
are observed.  This implies that the Crab had no stellar
companion to even stricter limits of $L \ltorder L_\odot$
and $M \ltorder M_\odot$ even if we are very conservative.

\section{Cassiopeia~A}

Figure~\ref{fig:image2} shows the co-added grizy PS1 image of a roughly 
2~arcmin region centered on Cas~A. The emission lines 
present in some of the bands show an outline of the remnant, and we have 
marked the geometric center (23:23:27.82, 58:48:49.4) of the remnant 
(\citealt{Thorstensen2001}) 
and the position of the neutron star (23:23:27.93, 58:48:42.5).
We adopt a distance of $3.4\pm0.3$~kpc (\citealt{Reed1995})
and an age of 330~years.  For this distance, a source
with velocity $100 v_2$~km/s will have moved $2\farcs0 v_2$ in the
330 years since the SN.  The 7\farcs0 distance of the NS from
the center of the SNR corresponds to a velocity of approximately
340~km/s.  

The PS1 extinction estimate at the distance of Cas~A 
is $E(B-V) \simeq 1.2$ ($A_V=3.7$) but it lies close to a sudden jump
in the extinction to $E(B-V) \simeq 1.5$ ($A_V=4.7$).  
This is more consistent with early estimates of $A_V \simeq 4.3$~mag by 
\cite{Searle1971} and lower than the estimates of $A_V \simeq 5.3$ 
to $6.2$~mag by \cite{Hurford1996} and $A_V = 6.2 \pm 0.6$~mag 
by \cite{Eriksen2009}. These estimates are based on 
using predicted and observed SNR emission line ratios to determine the
extinction. The distances used in the PS1 extinction
estimates are only reliable out to Cas~A.  

We used a generous selection radius of $30\farcs0$ from either the
center of the SNR or the NS, which corresponds to a velocity of
almost 1500~km/s.  These regions contain 15 PS1 sources of which
two are artifacts. This leaves thirteen stars, which we have again
labeled in order of their distance from the center of the SNR as shown
in Figure~\ref{fig:image2} and reported in Table~\ref{tab:casa1}.  
None of the stars have proper motion estimates in NOMAD and
three (\#4, \#9 and \#13) have proper motions in HSOY.  The 
predicted positions of these three stars at the time of the SN
are within a few arcseconds of their present positions and are
too small to display in Figure~\ref{fig:image2}.  They cannot
have been associated with the SN.  
 
The closest star (\#1) lies 11\farcs5 from the
center of the SNR and 16\farcs7 from the NS, corresponding to
velocities of $560$ and $820$~km/s that are not physical for 
the companion of a Type~II (IIb) SN.  The NS, while an
X-ray source, is not detected to very deep optical/near-IR
limits ($\gtorder 28$~mag at R band, \citealt{Fesen2006}).  
We again replaced the PS1 magnitudes with the results of
aperture photometry.  Here nebulosity is not an issue and
the PS1 PSF photometry produces good fits.  However, 
replacing the PS1 PSF photometry with forced aperture
aperture photometry allowed us to include photometry in
more bands than PS1 reports due to limits on 
signal-to-noise ratios.

Figure~\ref{fig:cmd2} shows the z/i$-$z CMD of the stars.
The brighter and closer (to the center) stars are
labeled.  Stars such as \#4, \#6 and \#9 which could have
$M > M_\odot$ at the distance of Cas~A (and assuming more 
extinction than the PS1 model) are at least $24\farcs0$ from
the center or the NS and would require $v \gtorder 1200$~km/s
to be associated with the SN.  Three of the four brightest
stars (\#4, \#9 and \#13) are also ruled out by their HSOY
proper motions.  There is no proper motion information for
\#6.   The two closest PS1 stars
would have to be $M \ltorder M_\odot$ and still require 
unreasonably high velocities. 

Table~\ref{tab:casa2} presents the results of fitting stellar
models to the SEDs.  Most of the stars can be well-fit, with
star \#5 as the worst case.  We again find that roughly half
the stars have properties consistent with the distance and
extinction of Cas~A.  Figure~\ref{fig:lum2} shows the 
luminosities and temperatures of the stars including the
extinction prior.  As suggested by the CMD, stars \#4,
\#6 and \#9 can have the highest luminosities, with \#4
potentially being a $\simeq 3 M_\odot$ B star with 
$L \simeq 10^{2.0}L_\odot$.  However, \#4 and \#9 also
have HSOY proper motions inconsistent with any association.  
The rest of the stars would
be low mass ($M < M_\odot$) dwarfs.  As with the Crab,
even these low mass stars are absent at distances from the
center corresponding to reasonable velocities, so it is
clear that any binary companion to Cas~A at death would have to have
$M \ltorder M_\odot$.

\section{SN~1987A}

\cite{Graves2005} obtained very tight limits on the presence of an optical
point source at the center of SN~1987A, with limits of
$\nu L_\nu < 0.26$, $0.33$, $0.15$, $0.30$ and $0.28 L_\odot$ in the
F330W, F435W, F555W, F625W and F814W filters.  To account
for dust absorption in their estimates, they note that the fraction of the
bolometric luminosity emerging in the infrared on day
2172 was 97\% (\citealt{Bouchet1993}) implying an 
effective optical depth of $\tau\simeq 3.5$. The effective optical
depth at the time their observations (day 6110) would then
be $\tau =0.45$ because of the $1/t^2$ dilution of the optical 
depth due to expansion.  This then implies that there cannot
be a binary companion (or other point source) more luminous
than roughly $L < 2 L_\odot$.  

\cite{Graves2005} discuss scenarios in 
which the dust might be clumped, but these scenarios were
based on clumpy dust distributions in a foreground screen
rather than a circumstellar medium. For a clumpy medium
closely surrounding the source, the extra absorption 
from a clump that happens to be along our line of sight 
will be partially balanced by the contribution from photons 
scattered onto our line of sight by other clumps (see
the discussion in \citealt{Kochanek2012b}). This
would essentially eliminate the worst case scenario
they consider, where there would be no dilution of the
optical depth by expansion and the luminosity constraint
would be $ \sim 30$ times weaker.  

A more important issue is that 
Herschel and ALMA observations imply the existence
of $M_d \sim 0.5$-$1.0M_\odot$ of dust in SN~1987A (\citealt{Matsuura2011}, 
\citealt{Indebetouw2014}, \citealt{Matsuura2015}), far more
than the amount inferred at early times as used by \cite{Graves2005}.
The characteristic visual optical depth scale for an SNR at time 
$t= 10 t_{10}$~years is 
\begin{equation} 
   \tau_0 = { 15 \kappa M_d M_e \over 64 \pi E t^2 }  
          \simeq 30 \kappa_4 M_{d0.1} M_{e10} E_{51}^{-1} t_{100}^{-2}.
\end{equation}
where a typical  dust visual opacity is $\kappa = 10^4 \kappa_4$~cm$^2$/g,
the dust mass is $M_d = 0.1 M_{d0.1} M_\odot$, the total 
ejected mass is $M_e = 10 M_{e10} M_\odot$ and the 
explosion energy is $E = 10^{51} E_{51}$~erg (see
\citealt{Kochanek2017}). At $t=16.7$~years (6110 days), the optical depth implied by the 
presently observed dust would have been
$\tau_0 \sim 10 \kappa_4 M_{d0.1} M_{e10} E_{51}^{-1}$
rather than the $\tau \simeq 0.5$ assumed by \cite{Graves2005}.
Even today, the optical depth would be of order 
$\tau_0 \sim 3 \kappa_4 M_{d0.1} M_{e10} E_{51}^{-1}$.
In short, given the amount of dust seen by Herschel and
ALMA, the \cite{Graves2005} observations provide no
useful limit on the luminosity of a binary companion
(or emission from any stellar remnant).

As a result, it is really only the mid-IR dust emission which
constrains the luminosity of any central source.  \cite{Matsuura2015}
find a total central dust luminosity in 2012 of $230 L_\odot$.
This luminosity is due to a combination of radioactive decay, absorption of
radiation from the expanding shocks, and any contribution from
a central source.  \cite{Matsuura2015} estimate that the available heating
from decay of $^{44}$Ti is $\simeq 400 L_\odot$, extrapolating
from \cite{Jerkstrand2011}, and that a further $\sim 50 L_\odot$
can be heating from the shocks exterior to the dusty region.
They do not discuss any additional contributions, but the 
two required heating sources already exceed the observed
luminosity, which suggests that only a modest fraction of the
observed $230 L_\odot$ could be due to a binary companion.
A reasonable upper limit is probably 10\% of the total 
luminosity, which corresponds to $M \ltorder 2.5 M_\odot$
(see Figure~\ref{fig:lum2}). A $3 M_\odot$ companion would
already represent half the observed luminosity.

\section{Discussion}

Examining the PS1 sources near the Crab and Cas~A, it is clear that these 
SN had no binary companion at death with a mass $M \gtorder M_\odot$. 
\cite{Graves2005} found a similar limit for SN~1987A, but 
the higher present day estimates of the dust content imply
a weaker limit of $M \ltorder 2.5 M_\odot$ from the observed
dust luminosity.
In terms of mass ratios, there are no companions to these
three ccSNe with initial mass ratios above $q \gtorder 0.1$. 
If the binary fraction at death is $f_b$, the probability of 
finding no binaries companions in three systems is $(1-f_b)^3$, implying 
that $f_b < 0.44$ at 90\% confidence.  

The stellar binary fraction of stars at death is generically lower than
that at birth. Even for passively evolving binaries with no 
interactions, only the explosions of the primary are in a 
stellar binary.  When the secondary explodes, the primary is
a compact object which is likely both difficult to detect and
need no longer be bound to the secondary. Following Equation~\ref{eqn:bfrac}
and again assuming a Salpeter IMF and a flat $P(q)$ distribution,
the factor $f_q=0.47$ if we detect all binaries
with $q>0.1$.  This implies that the initial binary
fraction is $F<0.61$ at 90\% confidence. 

There is no simple way to estimate the initial binary fraction
including binary interactions short of a full simulation of 
binary evolution, which is beyond our present scope. For example, 
the stars can merge, which
would reduce the numbers of binaries at death. Alternatively,
a primary with too little mass to explode can accrete enough
mass from the secondary to explode, which would tend to increase 
the numbers of binaries at death.  For example, if fraction
$f_M$ of binaries merge prior to the explosion of the primary,
then we really have the limit $ F (1-f_M) < 0.61$.  \cite{Sana2012}
argue that $f_M \simeq 0.25$, which would shift the limit
upwards to $F \ltorder 0.81$.  In their review of binary populations, \cite{Duchene2013}
cite a multiplicity frequency (fraction of multiple systems) with $q>0.1$ 
of $>60\%$ and $80\%$ for $8$-$16M_\odot$ and $>16 M_\odot$, respectively,
while \cite{Moe2016} find that only $16\pm9\%$ ($6\pm6)\%$ of $9$-$16M_\odot$  
($>16 M_\odot$) primaries are single. Thus, if there are only singles
or binaries, the binary fractions can be reconciled by a having a 
significant merger fraction.  

However, many massive stars are in higher order systems (triples, etc.),
and this reintroduces the tension.  
The fraction of massive stars in higher order systems is
high -- \cite{Moe2016} find
that $(52 \pm 13)\%$ ($73\pm 16$\%) of $9$-$16M_\odot$ ($>16 M_\odot$)
primaries are in higher order systems with $q>0.1$.  Let $f_H \simeq 62\%$ be the
fraction of non-single stars in higher order systems (i.e. 32\% are in binaries
and 52\% are in higher order systems, and $62=52/(52+32)$) .  The fraction
of exploding primaries is still $f_p = F/(1+F f_q)$ where $F$ is now 
the fraction of non-single stars at birth.  The fraction of
these with a stellar secondary and no additional companion is 
$(1-f_H)(1- f_M)$ where we allow fraction $f_M$ of binaries to merge.  
The fraction where there is an additional companion, so that there is
a stellar companion independent of whether the primary and secondary
have merged, is $f_H$.  Thus, the fraction of exploding primaries
with a stellar companion (not necessarily the original secondary) is
reduced by $(1-f_H)(1-f_M) + f_H \simeq 91\%$ rather than
$1-f_M \simeq 75\%$ for $f_H=62\%$ and a merger
fraction of $f_M = 25\%$.  Accounting for these higher order systems,
the limit on the initial fraction of non-single stars is
$F(1 - f_M + f_H f_M) < 0.61)$, leading to a limit of $ F \ltorder 0.67$ that
is again in significant tension with estimates of stellar 
multiplicities.  

This ignores any contribution from secondaries exploding in higher 
order systems which survive the explosion of the primary.  The fraction of exploding 
secondaries in this passive evolution model is 
$f_s = F f_q/(1+F f_q) \simeq 26\% $ for $f_q=0.47$ and $F=0.75$. 
In the absence of higher order systems, none of these would have 
stellar companions at death since the primary becomes a compact object
independent of whether the binary survives. With higher order systems,
fraction $f_H$ of the explosions of secondaries could also have
a stellar companion at death, although probably only a small 
fraction of these systems remain bound following the explosion of
the primary.

A second interesting point is that Cas~A was a Type~IIb SN 
(\citealt{Krause2008}, \citealt{Rest2008}) and yet
it cannot have been a binary at death unless the companion
was a dwarf star or a compact object.  
Many models for Type~IIb
SN invoke binary evolution and a massive companion, as was originally 
suggested by \cite{Podsiadlowski1993} and \cite{Woosley1994}
to explain SN~1993J.  Searches for an optical counterpart
to the Cas~A NS noted that any bound binary companion would have
to be very low mass (e.g., \citealt{Chakrabarty2001}, 
\citealt{Fesen2006}), but we could find no quantitative
discussion of limits on unbound binary companions. 
Cas~A is one of the SNR without candidate O/B star
companions in \cite{Guseinov2005} and the issue is 
mentioned in passing by \cite{Claeys2011}. In their
survey of possible binary models for Type~IIb SNe, \cite{Claeys2011}
essentially always were left with a companion that should have
been easily visible.  This suggests that caution should be 
exercised about invoking this theoretical motivation in searches
for companions to other Type~IIb SNe like SN~1993J (e.g.,
\citealt{Maund2004}, \citealt{Fox2014}) or SN~2011dh (e.g.,
\citealt{Folatelli2014}, \citealt{Maund2015}).       

That the absence of binary companions in only three systems already
has interesting implications suggests greatly expanding such searches.
The expected rate of companions is so high that it already surprising
not to have found one in these three systems, so even doubling the 
sample should either yield an example or indicate a
serious problem -- with six systems and no detection $f_b<28\%$
at 90\% confidence.  The three systems we consider here are 
a peculiar if well-defined sample, and it is not clear how
to incorporate the possible detection in SNR~S147 (\citealt{Dincel2015}).
If we simply treat it like a sample of four objects with one
detection, than $f_b \simeq 31\%$ with a symmetric 90\% 
confidence range of $8\% < f_b < 66\%$ that leaves much of 
tension intact.

The Crab and Cas~A are the easiest Galactic systems to examine due
to their youth and low extinction and stellar densities.  Searches
to binary companions of ccSNe are intrinsically easier than those
for Type~Ia because there is no immediate need to rule out the 
existence of faint, dwarf companions.  The searches should probably
not be limited to O/B stars (as in \citealt{Guseinov2005}).  For 
passively evolving binaries, companions will be dominated by
main sequence stars (\citealt{Kochanek2009}), but binary evolution 
and mass transfer greatly
broadens the spectrum of possible secondaries.  It is necessary
to separate the ccSNe from the Type~Ia's using the presence of 
a NS, the composition/structure of the SNR (e.g., \citealt{Lopez2009}, 
\citealt{Yamaguchi2014}), or, as done for Cas~A, light echo spectra.
Modern, multiband photometry such as the PS1 data used here, is an
important improvement, but studying large numbers of additional 
Galactic SNRs will only become relatively straight forward with the release of 
the full Gaia proper motion and parallax catalogs (e.g., \citealt{Gaia2016}).
Searching SNRs in the Magellanic Clouds is also feasible starting
from catalogs like \cite{Harris2004} or \cite{Nidever2017}.  While 
proper motions will be lacking, the fixed distance and modest
extinctions otherwise simplify the problem.

Identifying these binary companions to ccSNe is important. Not only
are they a key observational constraint on the role of binaries in
ccSNe, but they also provide important constraints on the formation
of runaway stars and the origins of NS velocities.  For example,
having the velocity of the former binary companion would greatly help to  
separate the contributions of binary disruption and explosive
kicks to the velocities of NS.  It would also be interesting to
identify such systems to observe the consequences of the SN 
explosion for the secondary.  This has primarily been considered
for stellar companions to Type~Ia SN (e.g., \citealt{Marietta2000},
\citealt{Shappee2013b}, \citealt{Pan2014}), but there should also
be long term effects on close companions to ccSNe.

\section*{Acknowledgments}

CSK thanks K. Auchettl, C. Badenes, S.~de~Mink, J.~J.~Eldridge, T. Holland-Ashford, L.~Lopez, K. Stanek, 
M. Pinsonneault, and T. Thompson for discussions and comments.   
CSK is supported by NSF grants AST-1515876 and
AST-1515927.  

\def\sp{\hphantom{-}}
\begin{table*}
   \centering
   \caption{Stars Near the Crab}
   \begin{tabular}{rrrrrccccc}
     \multicolumn{1}{c}{\#}
    &\multicolumn{1}{c}{$d_c$}
    &\multicolumn{1}{c}{$d_{NS}$}
    &\multicolumn{1}{c}{RA}
    &\multicolumn{1}{c}{Dec}
    &\multicolumn{1}{c}{g}
    &\multicolumn{1}{c}{r}
    &\multicolumn{1}{c}{i}
    &\multicolumn{1}{c}{z}
    &\multicolumn{1}{c}{y} \\
   \hline
       1 &12\farcs1 &10\farcs8 &83.633751 &22.011540 &$17.875\pm0.007$ &$16.749\pm0.005$ &$16.508\pm0.005$ &$16.232\pm0.005$ &$16.098\pm0.005$ \\
 2 &12\farcs2 & 5\farcs6 &83.634080 &22.015580 &$16.406\pm0.005$ &$15.429\pm0.004$ &$15.102\pm0.004$ &$14.941\pm0.003$ &$14.875\pm0.004$ \\
 3 &13\farcs2 & 5\farcs6 &83.632910 &22.012900 &$20.213\pm0.041$ &$18.479\pm0.018$ &$17.886\pm0.017$ &$17.490\pm0.009$ &$17.114\pm0.011$ \\
 4 &13\farcs2 &22\farcs0 &83.639220 &22.016270 &$17.145\pm0.005$ &$16.076\pm0.004$ &$15.767\pm0.003$ &$15.870\pm0.005$ &$15.513\pm0.004$ \\
 5 &13\farcs4 & 0\farcs5 &83.633031 &22.014530 &$17.207\pm0.009$ &$16.304\pm0.009$ &$16.088\pm0.009$ &$15.821\pm0.007$ &$15.687\pm0.009$ \\
 6 &14\farcs2 &12\farcs4 &83.633381 &22.011031 &$18.836\pm0.011$ &$17.822\pm0.011$ &$17.578\pm0.009$ &$17.343\pm0.009$ &$17.163\pm0.011$ \\
 7 &21\farcs6 &32\farcs2 &83.642270 &22.016590 &$21.955\pm0.330$ &$20.354\pm0.110$ &$18.801\pm0.029$ &$18.147\pm0.021$ &$17.785\pm0.025$ \\
 8 &28\farcs0 &41\farcs3 &83.643901 &22.009151 &$21.057\pm0.118$ &$19.822\pm0.053$ &$19.412\pm0.050$ &$19.336\pm0.057$ &$19.104\pm0.099$ \\
 9 &29\farcs1 &31\farcs3 &83.633680 &22.005790 &$22.833\pm0.327$ &$20.526\pm0.067$ &$19.380\pm0.044$ &$19.003\pm0.030$ &$18.544\pm0.035$ \\
10 &30\farcs3 &43\farcs4 &83.644091 &22.008290 &$19.849\pm0.030$ &$18.572\pm0.026$ &$18.195\pm0.013$ &$18.019\pm0.013$ &$17.729\pm0.022$ \\
11 &31\farcs3 &24\farcs3 &83.632790 &22.021190 &$18.055\pm0.009$ &$16.971\pm0.005$ &$16.813\pm0.009$ &$16.396\pm0.005$ &$16.398\pm0.007$ \\
12 &32\farcs9 &24\farcs8 &83.627980 &22.009300 &$20.487\pm0.066$ &$19.226\pm0.026$ &$18.745\pm0.019$ &$18.357\pm0.018$ &$18.355\pm0.059$ \\
13 &40\farcs1 &53\farcs7 &83.648400 &22.010360 &$15.976\pm0.004$ &$14.929\pm0.003$ &$14.580\pm0.003$ &$14.377\pm0.002$ &$14.258\pm0.003$ \\
14 &42\farcs1 &45\farcs7 &83.634560 &22.001831 &$21.708\pm0.131$ &$20.549\pm0.103$ &$21.173\pm0.168$ &$21.441\pm0.346$ &$20.392\pm0.264$ \\
15 &43\farcs8 &57\farcs1 &83.649970 &22.013350 &$19.691\pm0.019$ &$18.806\pm0.025$ &$18.246\pm0.011$ &$18.137\pm0.015$ &$17.996\pm0.030$ \\
16 &44\farcs3 &56\farcs2 &83.645360 &22.003920 &$19.306\pm0.025$ &$17.970\pm0.018$ &$17.516\pm0.009$ &$17.264\pm0.013$ &$17.008\pm0.018$ \\
17 &44\farcs5 &31\farcs4 &83.623531 &22.014000 &$21.403\pm0.155$ &$19.944\pm0.095$ &$19.034\pm0.046$ &$18.747\pm0.054$ &$18.441\pm0.064$ \\
18 &46\farcs3 &36\farcs4 &83.623841 &22.008830 &$19.301\pm0.043$ &$18.178\pm0.028$ &$17.915\pm0.022$ &$17.711\pm0.013$ &$17.572\pm0.028$ \\
19 &52\farcs5 &47\farcs0 &83.633171 &22.027510 &$21.674\pm0.130$ &$19.710\pm0.029$ &$18.614\pm0.024$ &$18.098\pm0.017$ &$17.837\pm0.021$ \\
20 &53\farcs9 &67\farcs1 &83.652970 &22.013090 &$18.549\pm0.013$ &$17.471\pm0.011$ &$16.953\pm0.009$ &$16.819\pm0.009$ &$16.612\pm0.013$ \\
21 &54\farcs1 &63\farcs6 &83.642910 &21.999390 &$17.840\pm0.011$ &$16.788\pm0.011$ &$16.540\pm0.007$ &$16.369\pm0.005$ &$16.098\pm0.009$ \\
22 &56\farcs5 &62\farcs2 &83.637380 &21.997660 &$16.915\pm0.005$ &$15.841\pm0.004$ &$15.490\pm0.005$ &$15.354\pm0.004$ &$15.119\pm0.005$ \\
23 &56\farcs7 &45\farcs2 &83.625051 &22.024660 &$21.395\pm0.129$ &$20.314\pm0.137$ &$19.727\pm0.071$ &$19.585\pm0.075$ &$19.326\pm0.088$ \\
24 &58\farcs8 &72\farcs4 &83.653070 &22.007001 &$19.107\pm0.013$ &$17.892\pm0.007$ &$17.595\pm0.009$ &$17.162\pm0.007$ &$17.182\pm0.015$ \\
25 &59\farcs3 &72\farcs2 &83.650590 &22.002890 &$20.473\pm0.041$ &$19.430\pm0.040$ &$19.089\pm0.036$ &$18.953\pm0.036$ &$18.766\pm0.054$ \\
26 &60\farcs6 &53\farcs2 &83.621881 &22.003800 &$19.146\pm0.015$ &$18.085\pm0.009$ &$17.933\pm0.009$ &$17.620\pm0.009$ &$17.480\pm0.013$ \\
27 &63\farcs8 &50\farcs2 &83.618310 &22.017740 &$19.334\pm0.018$ &$18.261\pm0.026$ &$17.996\pm0.015$ &$17.825\pm0.015$ &$17.564\pm0.021$ \\
28 &66\farcs5 &55\farcs4 &83.617581 &22.008580 &$17.723\pm0.007$ &$16.672\pm0.005$ &$16.724\pm0.007$ &$16.229\pm0.005$ &$16.051\pm0.007$ \\
29 &68\farcs1 &55\farcs2 &83.620091 &22.024130 &$21.746\pm0.236$ &$19.173\pm0.086$ &$18.695\pm0.018$ &$18.258\pm0.015$ &$17.765\pm0.030$ \\
30 &68\farcs3 &55\farcs9 &83.616450 &22.011591 &$18.793\pm0.017$ &$18.011\pm0.018$ &$17.899\pm0.011$ &$17.694\pm0.009$ &$17.448\pm0.017$ \\

   \hline
   \multicolumn{10}{l} {
   The stars are numbered in order of their distance from the center of the SNR ($d_c$) and the
distance from the NS ($d_{NS}$) is also given. } \\
   \multicolumn{10}{l} {
   The magnitudes are aperture magnitudes 
found from the PS1 images and an entry of $-$ indicates no detection. }\\
   \end{tabular}
   \label{tab:crab1}
\end{table*}

\begin{table*}
   \centering
   \caption{Fits to Stars Near the Crab}
   \begin{tabular}{rrrrcccc}
     \multicolumn{1}{c}{\#}
    &\multicolumn{1}{c}{$\chi^2_0$}
    &\multicolumn{1}{c}{$\chi^2_1$}
    &\multicolumn{1}{c}{$\chi^2_2$}
    &\multicolumn{1}{c}{$\log L_1/L_\odot$}
    &\multicolumn{1}{c}{$\log T_1 $}
    &\multicolumn{1}{c}{$\log L_2/L_\odot$}
    &\multicolumn{1}{c}{$\log T_2 $} \\
   \hline
       1 & 5.5 & 6.0 & 8.1 &$\sp 0.06\pm 0.25$ &$ 3.672\pm 0.030$ &$\sp 0.17\pm 0.30$ &$ 3.711\pm 0.004$ \\
 2 & 0.1 & 0.9 & 3.9 &$\sp 0.44\pm 0.24$ &$ 3.681\pm 0.024$ &$\sp 0.67\pm 0.20$ &$ 3.736\pm 0.004$ \\
 3 & 9.8 &10.3 &26.6 &$-0.11\pm 0.33$ &$ 3.676\pm 0.037$ &$-0.65\pm 0.47$ &$ 3.602\pm 0.005$ \\
 4 &21.9 &23.6 &30.7 &$\sp 0.28\pm 0.48$ &$ 3.716\pm 0.074$ &$\sp 0.27\pm 0.49$ &$ 3.740\pm 0.009$ \\
 5 & 3.6 & 5.6 & 5.7 &$\sp 0.40\pm 0.32$ &$ 3.764\pm 0.067$ &$\sp 0.35\pm 0.26$ &$ 3.745\pm 0.004$ \\
 6 &24.9 &30.9 &31.0 &$\sp 0.04\pm 0.55$ &$ 3.748\pm 0.077$ &$-0.17\pm 0.28$ &$ 3.730\pm 0.008$ \\
 7 & 0.1 & 0.5 & 0.7 &$-0.57\pm 0.23$ &$ 3.545\pm 0.021$ &$-0.67\pm 0.19$ &$ 3.533\pm 0.003$ \\
 8 & 5.1 & 9.5 &16.7 &$-0.71\pm 0.17$ &$ 3.659\pm 0.021$ &$-0.37\pm 0.12$ &$ 3.716\pm 0.008$ \\
 9 &13.5 &14.3 &17.8 &$-0.41\pm 0.47$ &$ 3.686\pm 0.069$ &$-0.97\pm 0.35$ &$ 3.566\pm 0.004$ \\
10 &17.8 &19.2 &19.9 &$-0.41\pm 0.37$ &$ 3.688\pm 0.046$ &$-0.50\pm 0.24$ &$ 3.681\pm 0.005$ \\
11 &12.7 &15.0 &16.8 &$\sp 0.04\pm 0.39$ &$ 3.709\pm 0.057$ &$-0.04\pm 0.38$ &$ 3.721\pm 0.006$ \\
12 & 8.2 &10.1 &10.5 &$-0.62\pm 0.30$ &$ 3.666\pm 0.040$ &$-0.61\pm 0.16$ &$ 3.668\pm 0.003$ \\
13 & 0.5 & 1.8 & 2.5 &$\sp 0.81\pm 0.23$ &$ 3.703\pm 0.020$ &$\sp 0.82\pm 0.18$ &$ 3.712\pm 0.003$ \\
14 &26.1 &54.1 &79.9 &$-0.29\pm 0.42$ &$ 3.714\pm 0.052$ &$\sp 0.05\pm 0.22$ &$ 3.762\pm 0.016$ \\
15 & 8.7 &10.0 &11.5 &$-0.37\pm 0.32$ &$ 3.706\pm 0.045$ &$-0.36\pm 0.11$ &$ 3.715\pm 0.004$ \\
16 & 3.6 & 3.8 & 3.9 &$-0.25\pm 0.23$ &$ 3.667\pm 0.025$ &$-0.23\pm 0.24$ &$ 3.662\pm 0.002$ \\
17 & 6.9 & 7.1 & 7.4 &$-0.71\pm 0.38$ &$ 3.650\pm 0.055$ &$-0.93\pm 0.14$ &$ 3.610\pm 0.002$ \\
18 &15.6 &18.0 &21.1 &$-0.40\pm 0.35$ &$ 3.691\pm 0.043$ &$-0.29\pm 0.19$ &$ 3.719\pm 0.006$ \\
19 &23.6 &27.6 &32.8 &$\sp 0.04\pm 0.64$ &$ 3.736\pm 0.096$ &$-0.83\pm 0.40$ &$ 3.564\pm 0.006$ \\
20 & 1.9 & 1.9 & 2.1 &$-0.11\pm 0.25$ &$ 3.685\pm 0.051$ &$-0.06\pm 0.23$ &$ 3.691\pm 0.003$ \\
21 & 9.1 & 9.5 &10.9 &$\sp 0.12\pm 0.35$ &$ 3.716\pm 0.063$ &$\sp 0.05\pm 0.36$ &$ 3.722\pm 0.005$ \\
22 & 3.7 & 4.3 & 5.7 &$\sp 0.39\pm 0.23$ &$ 3.682\pm 0.027$ &$\sp 0.56\pm 0.21$ &$ 3.710\pm 0.003$ \\
23 & 2.0 & 4.9 & 8.4 &$-0.89\pm 0.19$ &$ 3.629\pm 0.024$ &$-0.63\pm 0.08$ &$ 3.676\pm 0.003$ \\
24 & 8.2 &11.9 &12.6 &$-0.29\pm 0.34$ &$ 3.686\pm 0.044$ &$-0.32\pm 0.31$ &$ 3.689\pm 0.005$ \\
25 & 3.2 & 6.0 &12.4 &$-0.70\pm 0.16$ &$ 3.662\pm 0.019$ &$-0.29\pm 0.11$ &$ 3.726\pm 0.007$ \\
26 &22.3 &26.1 &27.4 &$-0.18\pm 0.46$ &$ 3.720\pm 0.059$ &$-0.20\pm 0.22$ &$ 3.731\pm 0.007$ \\
27 &17.7 &20.3 &21.6 &$-0.29\pm 0.40$ &$ 3.709\pm 0.051$ &$-0.28\pm 0.20$ &$ 3.721\pm 0.007$ \\
28 &37.2 &46.4 &46.9 &$\sp 0.70\pm 0.78$ &$ 3.828\pm 0.135$ &$\sp 0.15\pm 0.50$ &$ 3.736\pm 0.012$ \\
29 &28.3 &28.6 &45.2 &$\sp 0.04\pm 0.59$ &$ 3.737\pm 0.088$ &$-0.87\pm 0.42$ &$ 3.592\pm 0.004$ \\
30 &18.1 &27.5 &29.9 &$\sp 0.07\pm 0.54$ &$ 3.765\pm 0.071$ &$\sp 0.22\pm 0.20$ &$ 3.785\pm 0.011$ \\

   \hline
   \multicolumn{8}{l} {
The ID numbers are the same as in Table~\ref{tab:crab1}.  The goodnesses of fit
$\chi^2_0$, $\chi^2_1$, and $\chi^2_2$ are } \\
   \multicolumn{8}{l} {
for fits with no prior, a prior for the distance to the SN, and a prior on both the} \\
   \multicolumn{8}{l} {
distance and the extinction.
The probability weighted mean luminosities and } \\
   \multicolumn{8}{l} {
  temperatures are reported for
the latter two models. }\\
   \end{tabular}
   \label{tab:crab2}
\end{table*}

\def\sp{\hphantom{-}}
\begin{table*}
   \centering
   \caption{Stars Near Cas~A }
   \begin{tabular}{rrrrrccccc}
     \multicolumn{1}{c}{\#}
    &\multicolumn{1}{c}{$d_c$}
    &\multicolumn{1}{c}{$d_{NS}$}
    &\multicolumn{1}{c}{RA}
    &\multicolumn{1}{c}{Dec}
    &\multicolumn{1}{c}{g}
    &\multicolumn{1}{c}{r}
    &\multicolumn{1}{c}{i}
    &\multicolumn{1}{c}{z}
    &\multicolumn{1}{c}{y} \\
   \hline
       1 &11\farcs7 &16\farcs9 &350.869920 &58.816130 &$26.968\pm6.896$ &$22.959\pm0.243$ &$21.610\pm0.085$ &$21.066\pm0.096$ &$20.460\pm0.127$ \\
 2 &16\farcs6 &21\farcs9 &350.871040 &58.817400 &$24.223\pm0.478$ &$22.431\pm0.153$ &$21.512\pm0.089$ &$20.907\pm0.093$ &$20.540\pm0.143$ \\
 3 &19\farcs8 &26\farcs7 &350.862510 &58.818970 &$--$             &$23.087\pm0.241$ &$21.694\pm0.102$ &$20.755\pm0.072$ &$20.069\pm0.086$ \\
 4 &23\farcs8 &25\farcs3 &350.853010 &58.813060 &$16.802\pm0.003$ &$15.775\pm0.003$ &$15.306\pm0.003$ &$14.939\pm0.003$ &$14.712\pm0.003$ \\
 5 &25\farcs3 &32\farcs2 &350.861801 &58.820450 &$23.675\pm0.290$ &$22.005\pm0.119$ &$20.701\pm0.051$ &$20.375\pm0.051$ &$19.764\pm0.066$ \\
 6 &28\farcs0 &32\farcs9 &350.852650 &58.817580 &$19.784\pm0.009$ &$18.349\pm0.007$ &$17.633\pm0.005$ &$17.278\pm0.007$ &$17.003\pm0.009$ \\
 7 &28\farcs3 &24\farcs7 &350.855350 &58.807990 &$26.995\pm4.898$ &$23.297\pm0.335$ &$22.169\pm0.126$ &$21.046\pm0.096$ &$20.221\pm0.110$ \\
 8 &28\farcs5 &32\farcs6 &350.851430 &58.816520 &$24.604\pm0.551$ &$22.601\pm0.194$ &$21.244\pm0.066$ &$20.224\pm0.050$ &$19.396\pm0.046$ \\
 9 &28\farcs5 &22\farcs0 &350.863360 &58.805910 &$18.290\pm0.004$ &$17.258\pm0.004$ &$16.596\pm0.003$ &$16.304\pm0.004$ &$16.072\pm0.004$ \\
10 &29\farcs6 &23\farcs1 &350.873351 &58.806511 &$23.511\pm0.237$ &$21.663\pm0.069$ &$20.273\pm0.026$ &$19.595\pm0.018$ &$19.317\pm0.038$ \\
11 &32\farcs4 &27\farcs1 &350.878290 &58.807510 &$24.849\pm0.694$ &$22.528\pm0.179$ &$21.322\pm0.074$ &$20.694\pm0.062$ &$20.674\pm0.186$ \\
12 &34\farcs9 &29\farcs4 &350.878871 &58.806840 &$23.770\pm0.284$ &$22.723\pm0.270$ &$20.867\pm0.043$ &$20.308\pm0.042$ &$19.909\pm0.076$ \\
13 &35\farcs5 &29\farcs2 &350.861641 &58.804091 &$20.880\pm0.018$ &$19.655\pm0.011$ &$18.956\pm0.007$ &$18.598\pm0.011$ &$18.278\pm0.015$ \\

   \hline
   \multicolumn{10}{l}{
The stars are numbered in order of their distance from the center of the SNR ($d_c$) and the
distance from the NS ($d_{NS}$) is also given. } \\
   \multicolumn{10}{l}{
The magnitudes are aperture magnitudes 
found from the PS1 images and an entry of $--$ indicates no detection.} \\
   \end{tabular}
   \label{tab:casa1}
\end{table*}

\begin{table*}
   \centering
   \caption{Fits to Stars Near Cas~A}
   \begin{tabular}{rrrrcccc}
     \multicolumn{1}{c}{\#}
    &\multicolumn{1}{c}{$\chi^2_0$}
    &\multicolumn{1}{c}{$\chi^2_1$}
    &\multicolumn{1}{c}{$\chi^2_2$}
    &\multicolumn{1}{c}{$\log L_1/L_\odot$}
    &\multicolumn{1}{c}{$\log T_1 $}
    &\multicolumn{1}{c}{$\log L_2/L_\odot$}
    &\multicolumn{1}{c}{$\log T_2 $} \\
   \hline
       1 & 1.5 & 2.0 & 2.1 &$-0.81\pm 0.30$ &$ 3.635\pm 0.052$ &$-0.91\pm 0.12$ &$ 3.619\pm 0.010$ \\
 2 & 0.2 & 1.3 & 3.2 &$-1.08\pm 0.25$ &$ 3.598\pm 0.032$ &$-0.70\pm 0.12$ &$ 3.664\pm 0.013$ \\
 3 & 0.7 & 2.2 & 5.9 &$-0.07\pm 0.22$ &$ 3.738\pm 0.051$ &$-0.75\pm 0.14$ &$ 3.547\pm 0.009$ \\
 4 & 0.6 & 1.5 & 1.6 &$\sp 2.26\pm 0.50$ &$ 4.074\pm 0.152$ &$\sp 2.04\pm 0.48$ &$ 3.992\pm 0.131$ \\
 5 &15.3 &17.4 &17.6 &$-0.61\pm 0.33$ &$ 3.660\pm 0.062$ &$-0.53\pm 0.23$ &$ 3.684\pm 0.032$ \\
 6 & 0.3 & 0.4 & 1.8 &$\sp 0.44\pm 0.17$ &$ 3.703\pm 0.041$ &$\sp 0.57\pm 0.16$ &$ 3.768\pm 0.030$ \\
 7 & 3.2 & 5.7 &10.9 &$-0.00\pm 0.28$ &$ 3.741\pm 0.061$ &$-0.85\pm 0.16$ &$ 3.522\pm 0.014$ \\
 8 & 8.7 &16.4 &30.6 &$\sp 0.50\pm 0.37$ &$ 3.806\pm 0.076$ &$-0.59\pm 0.21$ &$ 3.526\pm 0.021$ \\
 9 & 0.2 & 0.8 & 0.9 &$\sp 1.28\pm 0.34$ &$ 3.913\pm 0.094$ &$\sp 1.26\pm 0.32$ &$ 3.906\pm 0.072$ \\
10 & 2.9 & 3.6 & 7.6 &$-0.65\pm 0.15$ &$ 3.558\pm 0.014$ &$-0.36\pm 0.17$ &$ 3.619\pm 0.015$ \\
11 & 2.8 & 3.4 & 4.1 &$-0.93\pm 0.29$ &$ 3.614\pm 0.048$ &$-0.78\pm 0.13$ &$ 3.645\pm 0.013$ \\
12 & 2.3 & 3.7 & 9.5 &$-1.06\pm 0.17$ &$ 3.523\pm 0.027$ &$-0.60\pm 0.21$ &$ 3.659\pm 0.031$ \\
13 & 4.1 & 7.2 & 9.4 &$\sp 0.06\pm 0.17$ &$ 3.770\pm 0.022$ &$-0.04\pm 0.13$ &$ 3.761\pm 0.011$ \\

   \hline
   \multicolumn{8}{l}{
The ID numbers are the same as in Table~\ref{tab:casa1}.  The goodnesses of fit
$\chi^2_0$, $\chi^2_1$, and $\chi^2_2$ are } \\
   \multicolumn{8}{l} {
for fits with no prior, a prior for the distance to the SN, and a prior on both the} \\
   \multicolumn{8}{l} {
distance and the extinction.
The probability weighted mean luminosities and } \\
   \multicolumn{8}{l} {
  temperatures are reported for
the latter two models. }\\
   \end{tabular}
   \label{tab:casa2}
\end{table*}

\end{document}